\documentclass[aps,pra,twocolumn,superscriptaddress,showpacs,10pt]{revtex4-1}
\usepackage{amsmath,amssymb,graphicx,color}
\usepackage{epstopdf}

\begin{document}

\title{Enhancing delocalization and entanglement in asymmetric discrete-time quantum walks}

\author{Hao Zhao}
\affiliation{College of Physics and Electronic Science, Hubei Normal University, Huangshi 435002, China}

\author{Qiyan He}
\altaffiliation{qiyanhe271@163.com}
\affiliation{College of Physics and Electronic Science, Hubei Normal University, Huangshi 435002, China}

\author{Fengzhi Yang}
\affiliation{College of Physics and Electronic Science, Hubei Normal University, Huangshi 435002, China}

\author{Cui Kong}
\affiliation{College of Physics and Electronic Science, Hubei Normal University, Huangshi 435002, China}
	
\author{Huiyun Cao}
\affiliation{College of Physics and Electronic Science, Hubei Normal University, Huangshi 435002, China}
	
\author{Tianqi Yan}
	\affiliation{College of Physics and Electronic Science, Hubei Normal University, Huangshi 435002, China}
	
	\author{Bingrui Zhong}
	\affiliation{College of Physics and Electronic Science, Hubei Normal University, Huangshi 435002, China}
	
	\author{Kaikun Tian}
	\affiliation{College of Physics and Electronic Science, Hubei Normal University, Huangshi 435002, China}
	
\author{Jiguo Wang}
\affiliation{College of Physics and Electronic Science, Hubei Normal University, Huangshi 435002, China}

\author{Chuanjia Shan}
\altaffiliation{cjshan@hbnu.edu.cn}
\affiliation{College of Physics and Electronic Science, Hubei Normal University, Huangshi 435002, China}
	
\author{Jibing Liu}
\altaffiliation{liujb@hbnu.edu.cn}
\affiliation{College of Physics and Electronic Science, Hubei Normal University, Huangshi 435002, China}
\affiliation{Hubei Engineering Institute, Huangshi 435003, China}

\begin{abstract}
 In this paper, we investigate the enhancement of delocalization and coin-position entanglement in asymmetric discrete-time quantum walks (DTQWs). The asymmetry results from asymmetric coin operations, asymmetric initial states, and asymmetric polarization-dependent losses. By varying these asymmetry factors, the inverse participation ratio and entanglement entropy of the walker are numerically calculated for different coin and loss parameters, both for symmetric and asymmetric initial states. We then experimentally implement a 16-step asymmetric DTQW using a time-multiplexing fiber loop structure. By choosing an asymmetric initial state, both coin-position entanglement and delocalization are simultaneously enhanced under specific coin parameters. Moreover, we observe that with finite asymmetric polarization-dependent loss, the photon probability on the left side decreases significantly, while that on the right side increases and becomes more localized. Interestingly, under specific coin parameters, the entanglement and delocalization exhibit improved robustness against polarization-dependent loss. These results demonstrate that the DTQWs constitute an ideal platform for investigating photonic delocalization and hybrid entanglement.
\end{abstract}

\maketitle
\section{Introduction}
Quantum walks are the quantum analogues of classical random walks\cite{aharonov1993quantum,travaglione2002implementing}. A key distinction is that quantum walks exhibit ballistic spreading due to quantum interference, whereas classical random walks exhibit Gaussian diffusion. Because of this ballistic spreading, quantum walks can achieve exponential speedup in hitting time compared to a classical random walk. This unique property has been utilised in quantum search algorithms\cite{shenvi2003quantum,potovcek2009optimized}, quantum computing\cite{childs2009universal,lovett2010universal,childs2013universal}, and quantum simulation\cite{schreiber20122d}. Quantum walks can generally be categorised as either continuous-time quantum walks (CTQWs) \cite{strauch2006connecting,mulken2011continuous} or discrete-time quantum walks (DTQWs)\cite{kempe2003quantum,tregenna2003controlling,schreiber20122d,schreiber2010photons,schreiber2011decoherence,venegas2012quantum}. In the DTQWs, the walker's movement direction is determined by the coin operation. The DTQWs provide a powerful platform for studying non-Hermitian phenomena, such as exceptional points\cite{xiao2021observation}, topological phase transitions\cite{kitagawa2010exploring,chen2018observation,lin2022topological}, and non-Hermitian skin effects \cite{xiao2020non,lin2022observation,cheng2025maximal}.\par 
An intrinsic property of DTQWs is the generation of hybrid entanglement between the internal (coin) and external (position) degrees of freedom of the walker during their evolution\cite{carneiro2005entanglement,abal2006quantum,schreiber20122d,cheng2025maximal}. The hybrid entanglement serves as a valuable quantum resource with significant applications in quantum information processing (QIP)\cite{barreiro2008beating,karimi2010spin,jeong2014generation,zhao2024robustness}. In ordered DTQWs, the entanglement entropy converges to a stable value as the number of steps increases. Notably, this value typically does not reach its maximum and is highly dependent on the initial state\cite{abal2006quantum}. Quantum entanglement is also highly susceptible to polarization-dependent losses, decoherence, and environmental noise. Another important phenomenon observed in quantum walk systems is localization\cite{inui2004localization,chandrashekar2015localized,giraud2005quantum,keating2007localization,schreiber2011decoherence,crespi2013anderson,buarque2019aperiodic}. Inhomogeneous environmental conditions can disrupt interference between different paths of a single particle, causing the wave packet to remain localized rather than spreading over the entire lattice. The localization of quantum walks has been experimentally studied for potential applications in quantum storage\cite{chandrashekar2015localized}. However, it may hinder quantum algorithmic applications that require faster spreading \cite{giraud2005quantum,keating2007localization}. The implementation of dynamical disorder in DTQWs has been shown to increase entanglement entropy while simultaneously inducing localization of the quantum walker\cite{vieira2013dynamically,fang2023maximal}. Therefore, it is crucial to explore methods that simultaneously enhance both delocalization and entanglement in quantum walk systems while improving their robustness against asymmetric polarization-dependent losses.\par 
The DTQWs have been implemented on various physical platforms, including nuclear magnetic resonance\cite{ryan2005experimental}, trapped ions\cite{karski2009quantum,schmitz2009quantum}, optical systems\cite{schreiber2010photons,schreiber2011decoherence}, and superconducting qubits\cite{ramasesh2017direct}. Among these, optical systems are a particularly significant platform for implementing DTQWs, due to their high coherence and scalability. Experimental realizations in optical systems are primarily based on time-multiplexing\cite{chen2018observation,lin2022topological,lin2022observation,cheng2025maximal} or spatial displacement schemes\cite{xue2015experimental,zhan2017detecting,xiao2021observation,xiao2020non}. Bulk optical setups based on spatial displacers are limited by large physical dimensions, restricted scalability, and insufficient stability, hindering the realization of long-step quantum walks. In contrast, time-multiplexed optical systems using fiber-loop configurations can achieve much longer walk steps by encoding different photon position states in the time domain. This approach enables precise control over the walker’s internal states and effectively enlarges the accessible position space. Furthermore, such systems can generate maximal coin-position entanglement via specific coin operations with a fixed initial state\cite{cheng2025maximal}. However, current research rarely explores the simultaneous enhancement of entanglement and delocalization while maintaining robustness against asymmetric losses in time-multiplexed DTQW platforms.\par 
We present a one-dimensional 16-step DTQW implemented using a time-multiplexed fiber-loop system. To investigate the quantum walk's properties, we analyze the inverse participation ratio and entanglement entropy as functions of the initial states, coin operations, and polarization-dependent losses. By introducing controllable loss and tunable coin parameters, the walker's probability distributions are numerically and experimentally analyzed. The results show that a specific coin operation and initial state can effectively enhance both entanglement and delocalization. Furthermore, for certain coin parameters, the system exhibits improved robustness of entanglement and delocalization against asymmetric losses. The ability to flexibly adjust loss and coin parameters within the fiber-loop DTQW platform offers new opportunities for robust quantum-state preparation.

\section{Theoretical analysis}
For a one-dimensional DTQW, the Hilbert space is $H=H_{C} \otimes H_{P} $. $H_{C}$ is a two-dimensional coin space spanned by the horizontal and vertical polarization states, $\left | H  \right \rangle $ and $\left | V  \right \rangle $, respectively. $H_{P} $ is an infinite-dimensional position space spanned by the orthogonal basis states $\left | x \right \rangle $, where each basis represents the walker’s position $x$ on a one-dimensional lattice site.
The initial state of the quantum walker is expressed as $\left | \psi_{0}   \right \rangle =\left [ \cos \phi  \left | H \right \rangle
+ i\sin \phi \left | V \right \rangle \right ]\left | 0  \right \rangle $. Here, $\phi \in \left [ 0,\pi  \right ] $ denotes the initial state parameter. \par 
In the DTQW, each step consists of a coin operation followed by a conditional shift operation determined by the coin state. Accordingly, the $t$-step evolution operator can be written as $\hat{U}(t)=\bigl[\hat{S}\hat{M}(\hat{C}\otimes \hat{I}_{p})\bigr]^{t}$. The coin operator is defined as $\hat{C} = \bigl(\begin{smallmatrix}
	\cos \theta & \sin \theta \\
	\sin \theta & -\cos \theta
\end{smallmatrix}\bigr)$, acting on the polarization state of the photon at position $x$ and time $t$. The parameter $\theta \in \left [ 0^{\circ},180^{\circ} \right ] $ denotes the coin parameter.  Here, $\hat{I}_{p}$ is the identity operator in the position space, which ensures that the coin operation acts solely on the polarization degree of freedom without affecting the walker’s position state. The loss operator is given by $\hat{M} = \sum_{x} |x\rangle\langle x|\left( |H\rangle\langle H| + e^{-\gamma} |V\rangle\langle V| \right)$, where $\gamma$ denotes the asymmetric loss between the H-polarized and V-polarized components. When $\gamma\neq0$, $\hat{M}$ is a non-unitary operator. The conditional shift operator is $\hat{S} =\sum_{x}^{} \left (| x-1\right\rangle \left\langle x\right| \otimes\left| V\right\rangle \left\langle V\right|
+ \left| x+1\right\rangle \left\langle x\right|\otimes\left| H\right\rangle \left\langle H\right|) $, which moves the walker one site to the left (right) if it is in the vertical (horizontal) polarization state. The coherent action of $\hat{S}$ and $\hat{C}$ generates coin–position entanglement between the walker’s polarization and position states. After $t$ steps, the walker's state is given by $\left | \psi_{x} \left ( t \right )  \right \rangle =\sum_{x}^{} \left [ a\left ( x,t \right ) \left | H \right \rangle
+ b\left ( x,t \right ) \left | V \right \rangle \right ]\left | x  \right \rangle  $, where $a\left ( x,t \right ) $ and $b\left ( x,t \right ) $ represent the probability amplitudes of the horizontally and vertically polarized states, respectively. These amplitudes satisfy the normalization condition $\sum_{x}^{} \left [ \left | a\left ( x,t \right )  \right |^{2} + \left | b\left ( x,t \right )  \right |^{2} \right ]=1 $. \par 
The probability of detecting the photon at position $x$ after $t$ steps is given by $P_{t} \left ( x \right ) =\left | a_{t} \left ( x \right )  \right |^{2} +\left | b_{t} \left ( x \right )  \right |^{2}$. The inverse participation ratio (IPR) of the probability distribution is commonly used to characterize the localization or delocalization of the walker\cite{giraud2005quantum,buarque2019aperiodic,naves2022enhancing}, and is defined as
\begin{equation}
	\mathrm{IPR} (t) =\sum_{x} \left [ P_{t} \left (x\right )   \right ] ^{2}. 
\end{equation}
When $\mathrm{IPR} (t) \approx 1$, the walker remains localized, whereas $\mathrm{IPR} (t) \approx 1/N$ indicates complete delocalization. Here, $N$ represents the maximal number of positions for the probability $P_{t} \left ( x \right ) $ distribution. By tracing out the position degrees of freedom, the reduced density matrix of the coin state is obtained as
\begin{equation}
	\rho _{c} =\mathsf{Tr_{p}} \left [ |\psi_{x} \left ( t \right )  \rangle\langle \psi_{x} \left ( t \right )  | \right ]=\left ( \begin{matrix}
		\alpha \left ( t \right ) &\chi  \left ( t \right )  \\
		\chi ^{\ast }  \left ( t \right ) &\beta \left ( t \right ) 
	\end{matrix} \right ),
\end{equation}
where the diagonal elements $\alpha \left ( t \right ) = {\textstyle \sum_{x}^{}} \left | a \left ( x,t \right )  \right |^{2} $ and $\beta  \left ( t \right ) = {\textstyle \sum_{x}^{}} \left | b \left (x, t \right )  \right |^{2} $, correspond to the probabilities that the coin state is in the horizontal and vertical states, respectively. The off-diagonal element $\chi  \left ( t \right ) = {\textstyle \sum_{x}^{}} a \left (x, t \right ) b^{\ast }\left ( x,t \right ) $ denotes the coherence between the two coin states. To quantify the coin–position entanglement, we use the von Neumann entropy\cite{carneiro2005entanglement,abal2006quantum,cheng2025maximal}, given by
\begin{equation}
	S_{\mathrm{ E}} \left ( \rho _{c}  \right )=-\lambda _{1} \log_{2}{\lambda _{1} } -\lambda _{2} \log_{2}{\lambda _{2} } ,
\end{equation}
where $\lambda _{1,2}$ are the eigenvalues of $\rho _{c}$, expressed as $\lambda _{1,2} =\frac{1\pm\sqrt{1-4\left [ \alpha \left ( t \right ) \beta \left ( t \right ) -\left | \chi \left ( t \right )  \right | ^{2}  \right ] }  }{2}$. This entanglement entropy quantifies coin-position entanglement after $t$ steps in the DTQW system.\par   
 \begin{figure}[ht!]
	\centering\includegraphics[width=6.8cm]{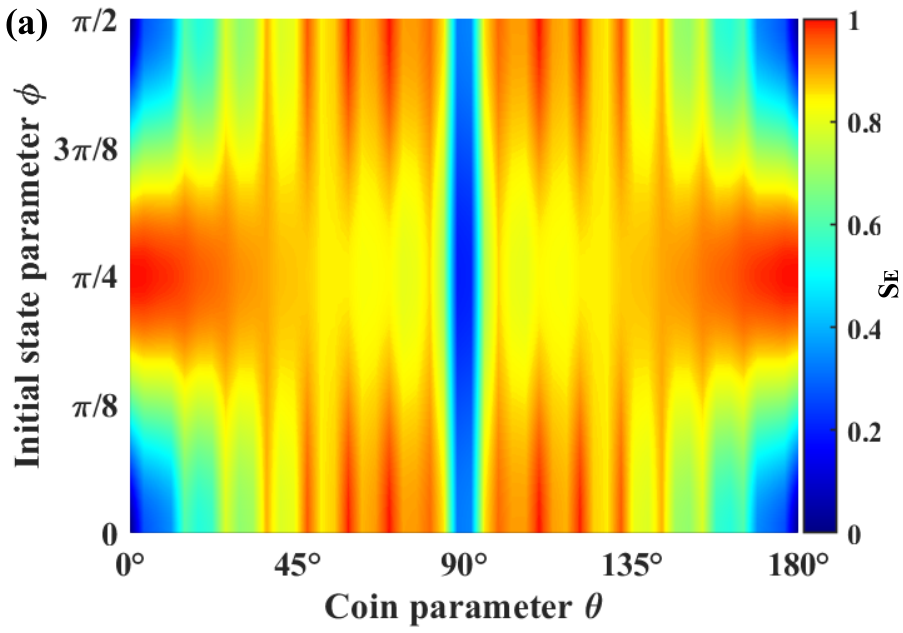}
	\centering\includegraphics[width=6.8cm]{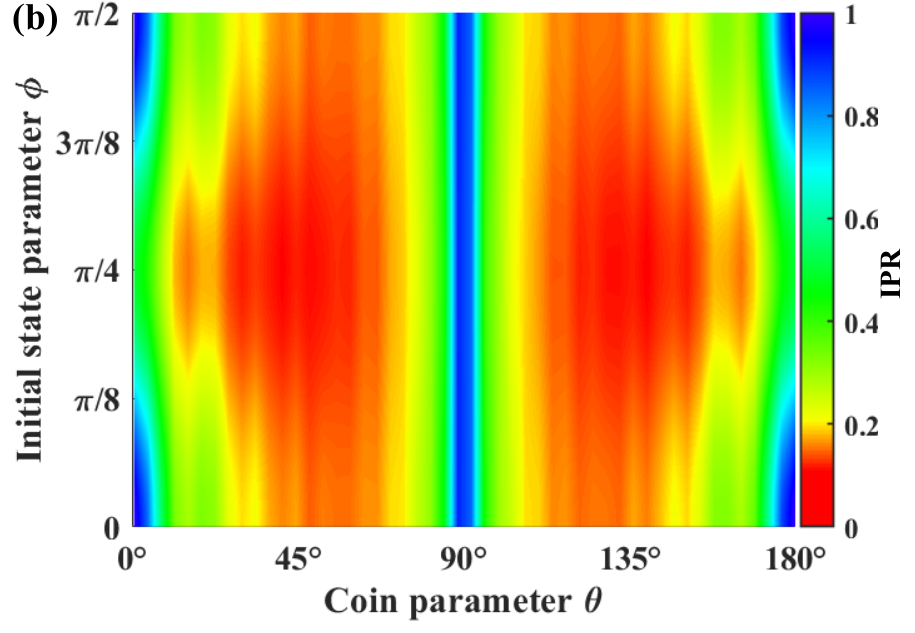}  
	\caption{(a) Coin-position entanglement entropy $S_{\mathrm{ E}}$ versus coin parameter $\theta$ and initial state parameter $\phi $. (b) IPR of the walker versus coin parameter $\theta$ and initial state parameter $\phi $.}
\end{figure}
Fig. 1(a) shows how the von Neumann entropy $S_{E}$ of a 16-step quantum walk varies with the coin parameter $\theta$ and the initial state parameter $\phi$, based on numerical calculations using Eq. (3). The loss parameter is fixed at $\gamma=0$. Since $S_{E}$ is symmetric with respect to $\theta = 90^\circ$, we restrict the coin parameter $\theta$ to the interval $(0^{\circ}, 90^{\circ})$. For $\phi = 0$, corresponding to the initial state $\lvert 0 \rangle \otimes \lvert H \rangle $, $S_{E}$ initially increases and subsequently decreases as $\theta$ increases, exhibiting local oscillations. When the walker’s von Neumann entropy $S_{E}$ exceeds 0.95, the coin parameter $\theta$ is confined to the ranges $47.3^{\circ}<\theta<48.7^{\circ}$, $56.9^{\circ}<\theta<60.8^{\circ}$, and $67.3^{\circ}<\theta<72^{\circ}$. This phenomenon arises from the asymmetric initial state and coin operation, which lead to unequal flipping probabilities between the horizontal ($\lvert H \rangle$) and vertical ($\lvert V \rangle$) polarization components, thereby disrupting the interference balance between them along different paths. Consequently, the entanglement entropy is enhanced within these narrow ranges of $\theta$. When the initial state parameter $\phi $ is set to $\pi/4$, corresponding to a symmetric initial state $|0 \rangle \otimes \left ( |H \rangle+i|V \rangle \right )  /\sqrt{2}$, the entanglement entropy $S_{E}$ decreases as $\theta$ increases. This behavior arises because the $\left | H \right \rangle $ component is partially converted into $\left | V \right \rangle $ and propagates toward the $ x-1$ direction, while the $\left | V \right \rangle $ component is partially converted into $\left | H \right \rangle $ and moves toward the $x+1$ direction during the walk. As a result, the two polarization components partially overlap at each position, which reduces the quantum correlation between the coin and position degrees of freedom. For the symmetric initial state, the walker’s von Neumann entropy $S_E$ exceeds 0.95 within the approximate range $0^{\circ } < \theta < 20.4^{\circ}$. Therefore, the entanglement entropy $S_E$ can reach near-maximum values only within a restricted range of the coin parameter in non-Hadamard DTQWs.\par 

\begin{figure}[ht!]
	\centering\includegraphics[width=6.8cm]{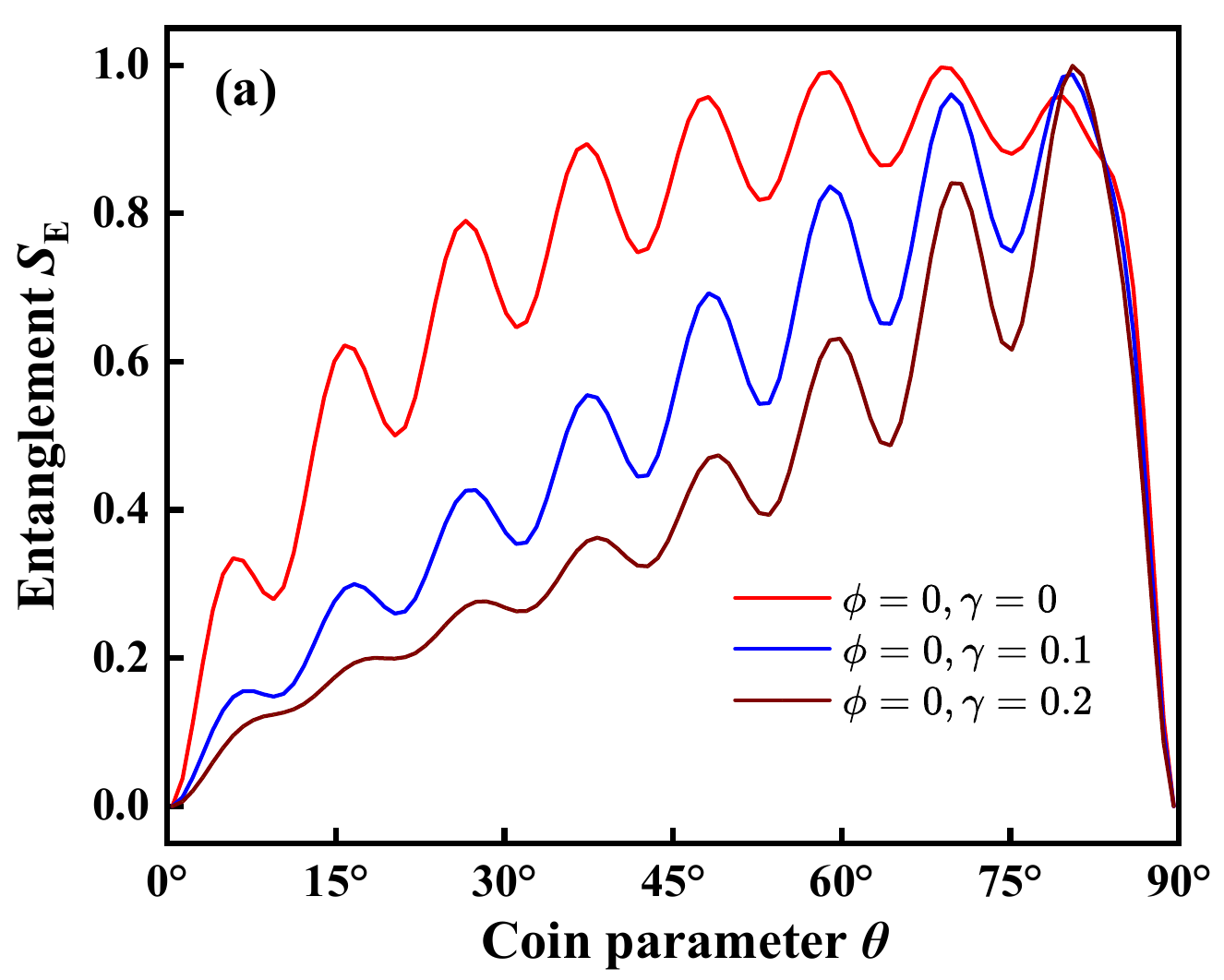}
	\centering\includegraphics[width=6.8cm]{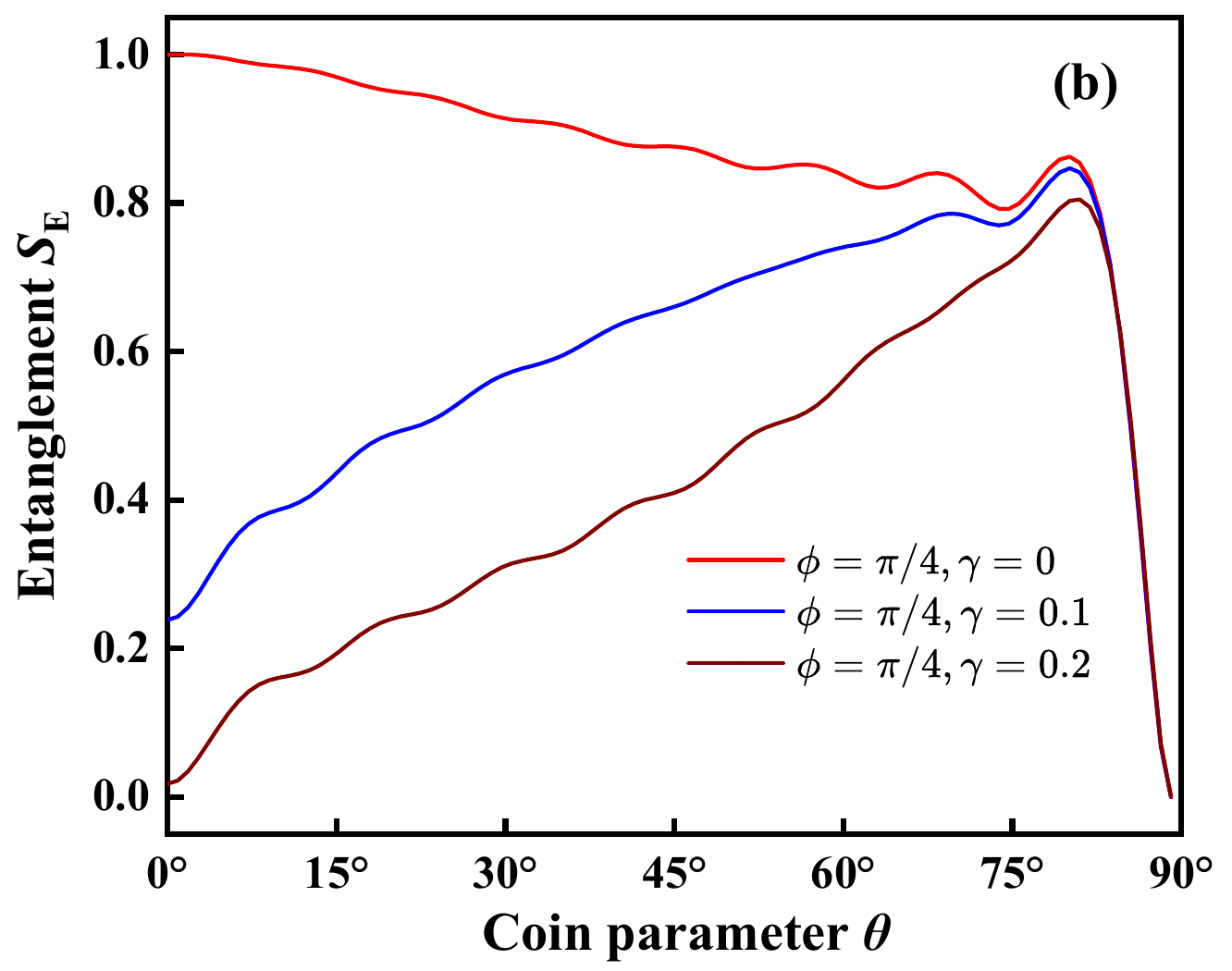}  
	\caption{(a) Coin-position entanglement entropy $S_{E}$ versus coin parameter $\theta $ and loss parameter $\gamma$. (a) Initial state parameter $\phi=0$. (b) Initial state parameter $\phi=\pi /4$. The red, blue, and wine lines correspond to numerical simulations with the loss parameter $\gamma=0$, $\gamma=0.1$, $\gamma=0.2$, respectively.}
\end{figure}
Fig. 1(b) shows how the IPR of a 16-step quantum walk varies with the coin parameter $\theta $ and the initial state parameter $\phi$, with the loss parameter fixed at $\gamma=0$, based on numerical calculations using Eq. (1). The coin parameter $\theta$ is considered within the interval $(0^{\circ}, 90^{\circ})$. For an arbitrary initial state parameter $\phi $, the $\mathrm{IPR }$ initially decreases and then increases as $\theta$ increases. This phenomenon occurs because setting $\theta = 0^{\circ}$ makes the coin operator an identity operation, so the unitary evolution $\hat{U}$ preserves the walker’s polarization. Walkers in the horizontal polarization state become fully localized at position $x=16$, whereas those in the vertical polarization state are localized at position $x=-16$, indicating that the walker is in a localized state. As $\theta$ increases from $0^{\circ}$, the coin operation gradually mixes the $\left | H \right \rangle $ and $\left | V \right \rangle $ components, causing the walker to spread over more positions, and leading to a decrease in the IPR. As the coin parameter increases further, the coin operator more effectively exchanges the $\lvert H \rangle$ and $\lvert V \rangle$ components, which suppresses the walker’s spatial spreading and leads to an increase in the IPR. Notably, when $\phi = 0$, the IPR drops below 0.18 within the range $36.9^{\circ} < \theta < 68.2^{\circ}$, while for $\phi = \pi/4$, the corresponding range is $21.9^{\circ} < \theta < 70.7^{\circ}$. As shown in Fig. 1(a), for a nearly symmetric initial state, the ranges of the coin parameter for higher entanglement ($E > 0.95$) and stronger delocalization ($\mathrm{IPR } < 0.18$) do not overlap. This means entanglement and delocalization cannot be optimized together for a symmetric initial state. In contrast, for asymmetric initial states, the ranges of coin parameter partially overlap, enabling simultaneous enhancement of both entanglement and delocalization.\par 

Fig. 2(a) shows the entanglement entropy $S_{E}$ of a 16-step quantum walk versus the coin parameter for various loss parameters, based on numerical calculations using Eq. (3). Here, we focus on the case $\phi = 0$. The entanglement entropy exhibits local peak values within several narrow ranges of the coin parameter, such as near $37^{\circ}$, $48^{\circ}$, and $59^{\circ}$. $S_{E}$ decreases gradually as the loss parameter increases, and this effect diminishes as the coin parameter increases. Fig. 2(b) presents the entanglement entropy $S_{E}$ of a 16-step quantum walk versus the coin parameter for various loss parameters, with the initial state fixed at $\phi =\pi/4 $. The entanglement entropy $S_{E}$ also decreases as the loss parameter increases. Moreover, as the coin parameter increases, the decrease of $S_{E}$ with respect to the loss parameter becomes slower. Therefore, with larger coin parameters, the entanglement entropy becomes more robust against the loss parameter $\gamma$. \par
\begin{figure}[ht!]
	\centering\includegraphics[width=6.8cm]{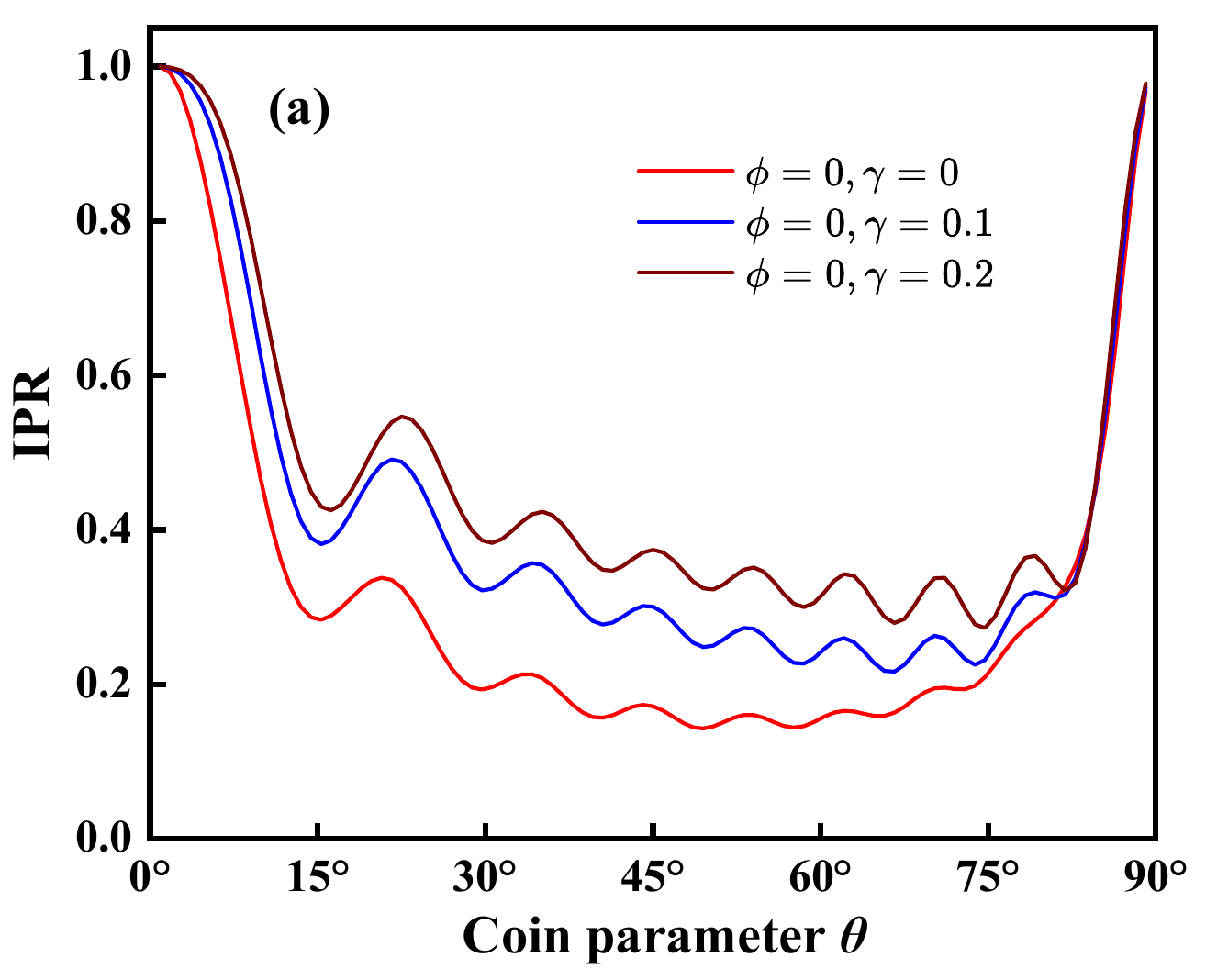}
	\centering\includegraphics[width=6.8cm]{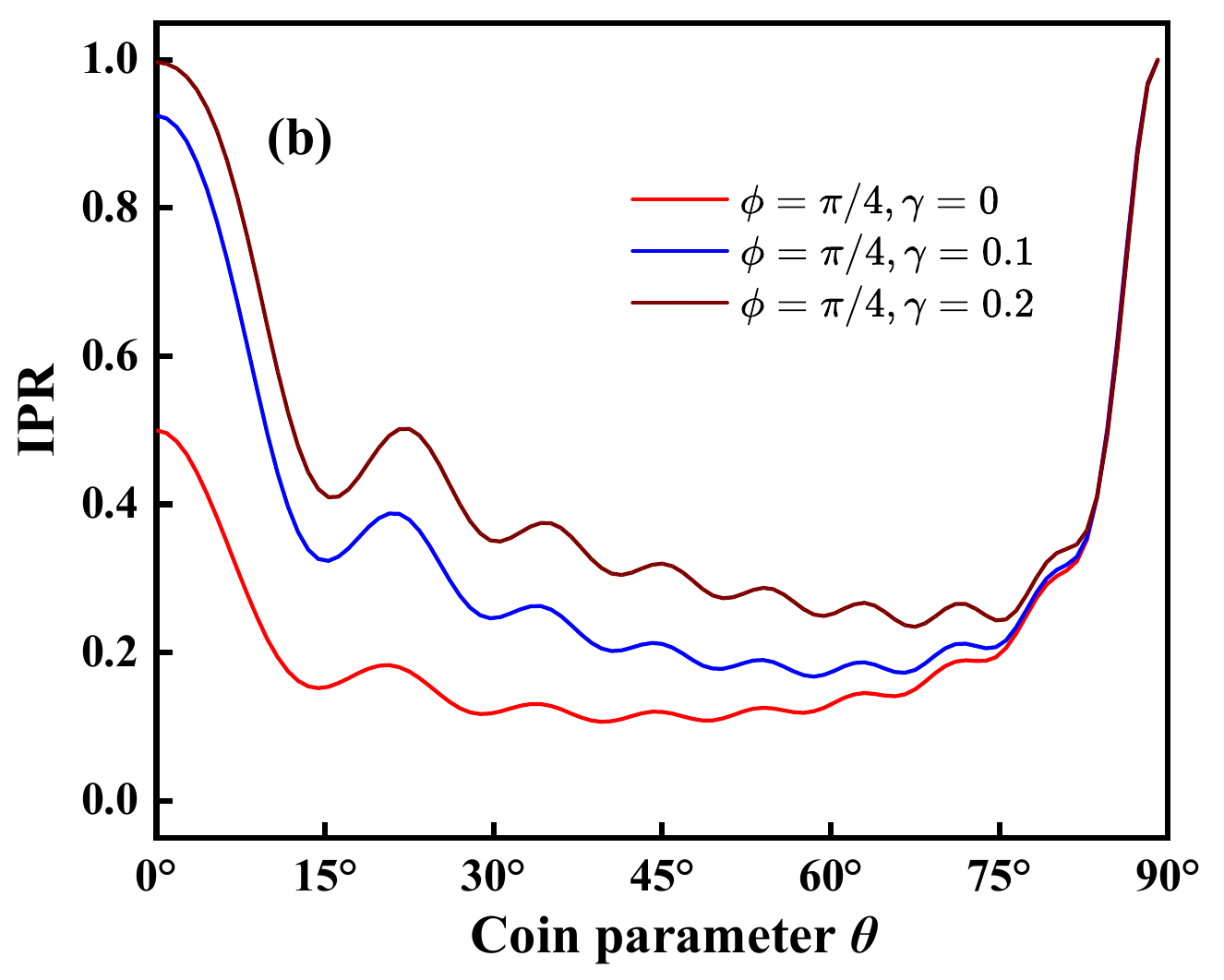}  
	\caption{(a) IPR of the walker versus coin parameter $\theta $ and loss parameter $\gamma$. (a) Initial state parameter $\phi=0$. (b) Initial state parameter $\phi=\pi /4$. The red, blue, and wine lines correspond to numerical simulations with the loss parameter $\gamma=0$, $\gamma=0.1$, $\gamma=0.2$, respectively.}
\end{figure}
Fig. 3(a) and 3(b) show the IPR of a 16-step quantum walk versus the coin parameter for various loss parameters, obtained from Eq. (1). The initial state parameter is set to $\phi=0$ in Fig. 3(a) and $\phi=\pi /4$ in Fig. 3(b). Comparison of Fig. 2(a) and Fig. 3(a) indicates that, for initial state parameter $\phi = 0$ and coin parameter $\theta = 0^{\circ}$, the entanglement entropy remains nearly zero, while the IPR approaches 1 regardless of the loss parameter. This result arises because the coin operation acts as the identity, resulting in the walker remaining localized at position $x = t$ and preventing the generation of quantum correlations between the coin and position degrees of freedom. In contrast, comparing Fig. 2(b) and Fig. 3(b) shows that, when the initial state parameter is $\pi /4$ and the coin parameter is $0^{\circ }$ with the loss parameter $\gamma = 0$, the entanglement entropy remains nearly 1, and the IPR approaches 0.5. In this scenario, the coin operator is the identity operator, causing the walker’s H-polarization component to occupy only the position $x = t$ and the V-polarization component to occupy only the position $x = -t$, with equal probability. This scenario produces a perfect correlation between the coin and position states. The loss operator $\hat{M}$ selectively attenuates the $\left | V \right \rangle $ component, reducing the probability of the walker being in the $\left | V \right \rangle $ state and increasing the probability of the $\left | H \right \rangle $ state. Therefore, as the loss parameter increases, coin-position entanglement is gradually suppressed, while the IPR correspondingly increases. Furthermore, regardless of the initial state parameter($\phi = 0$ or $\phi = \pi/4$), the IPR in both cases exhibits an overall decreasing trend followed by an increasing trend as the coin parameter $\theta$ varies from $0^{\circ}$ to $90^{\circ}$. Importantly, for an asymmetric initial state, as shown in Fig. 2(a) and Fig. 3(a), entanglement and delocalization of walker can be simultaneously optimized by appropriately selecting the coin parameter. \par

\section{Experimental demonstrations}
 \begin{figure*}[ht!]
	\centering\includegraphics[width=13cm]{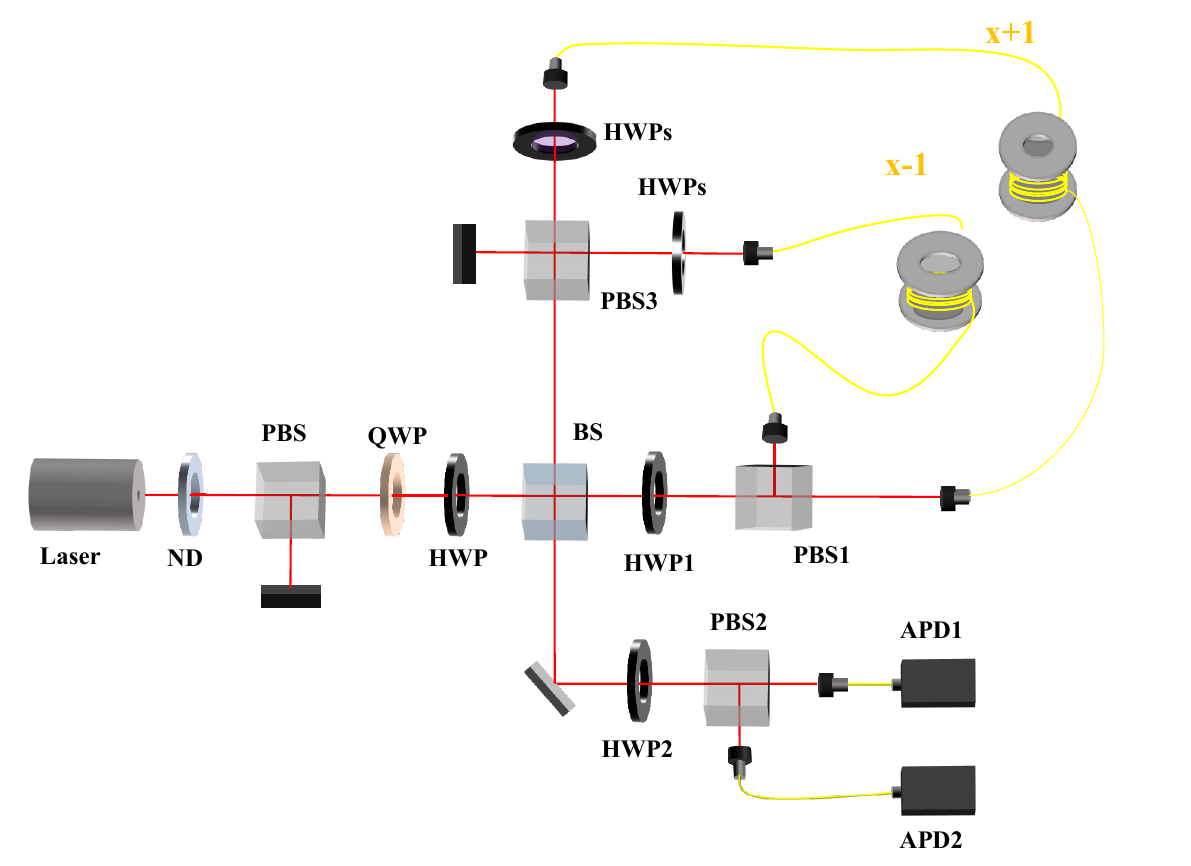} 
		\begin{minipage}{1\textwidth}
		\caption{Schematic of the experimental setup. The laser beam is attenuated to single-photon levels by a neutral-density  (ND) filter and coupled into the loop via a 90/10 beam splitter (BS). HWP: half-wave plate; QWP: quarter-wave plate; PBS: polarization beam splitter; APD: single-photon avalanche photodetector.}
	\end{minipage} 
\end{figure*}
As shown in Fig. 4, a photon wave packet is generated by a pulsed laser with a central wavelength of 810 nm, a pulse width of 80 ps, and a repetition rate of 125 kHz. The pulse intensity is attenuated to the single-photon level using a neutral-density (ND) filter. The initial state is prepared with a standard half-wave plate (HWP) and quarter-wave plate (QWP). The quantum coin operation is realized by HWP1. The shift operation is implemented using an optical feedback loop. Horizontally and vertically polarized photons are spatially separated from a polarization beam splitter (PBS1). Horizontally polarized photons propagate through the long fiber loop and experience a 155 ns delay. Vertically polarized photons travel through the short fiber loop with a 150 ns delay. This 5 ns difference ensures that the two polarization components are deterministically separated in time. At the output, the two paths are coherently recombined, and the photons are directed back to the input beam splitter (BS) to couple into the optical fiber loop. After one full evolution, the photon wave packet is distributed across several time windows, each corresponding to a discrete spatial position. The probability for an H(V)-polarized photon to complete a full round trip through the long (short) fiber loop is approximately 0.58. To measure the photon’s probability distribution without disturbing its evolution in the loop, only about 10$\%$ of the photons are coupled out of the loop through the BS. The detection setup consists of an HWP2, PBS2, and two avalanche photodiodes (APDs). These components are used to measure the temporal and polarization properties of the photons.\par
Due to the polarization-dependent losses existing in the fiber loop, we introduce a loss parameter $\gamma$ to characterize the relative attenuation between the H- and V-polarized photons. The specific details for controlling the loss parameter $\gamma$ are as follows. When the losses for H-polarized and V-polarized photons in the fiber loop are balanced by properly adjusting the angle of the HWPs, the number of either H- or V-polarized photons injected into the fiber loops within a certain time window is $N_0$. After one round-trip step ($t = 1$), the number of remaining photons is $ N_{0} \times 0.58 $. The loss operator $ \hat{L} \left ( \gamma  \right )$ leaves the H-polarized photons unaffected but introduces a tunable attenuation for V-polarized photons. Therefore, for a nonzero value of $\gamma$, the number of V-polarized photons remaining after one round-trip becomes $N_{0} \times 0.61\times e^{-2\gamma }$. Here, $\gamma $ characterizes the polarization-dependent loss. By tuning the HWPs in the short fiber loop, a fraction of the V-polarized photons is directed out through PBS3, thereby enabling a controllable loss parameter $\gamma$. To characterize the coin-position state, projection measurements are performed in the polarization bases $\{\lvert H\rangle, \lvert V\rangle\}$ and $\{\tfrac{1}{\sqrt{2}}(\lvert H\rangle + \lvert V\rangle),\, \tfrac{1}{\sqrt{2}}(\lvert H\rangle - \lvert V\rangle)\}$ to experimentally reconstruct the reduced density matrix $\rho_{c}$. The eigenvalues of $\rho_c$ are then obtained to quantify the entanglement entropy between the coin and position state.\par
 \begin{figure*}[ht!]
	\centering\includegraphics[width=6.4cm]{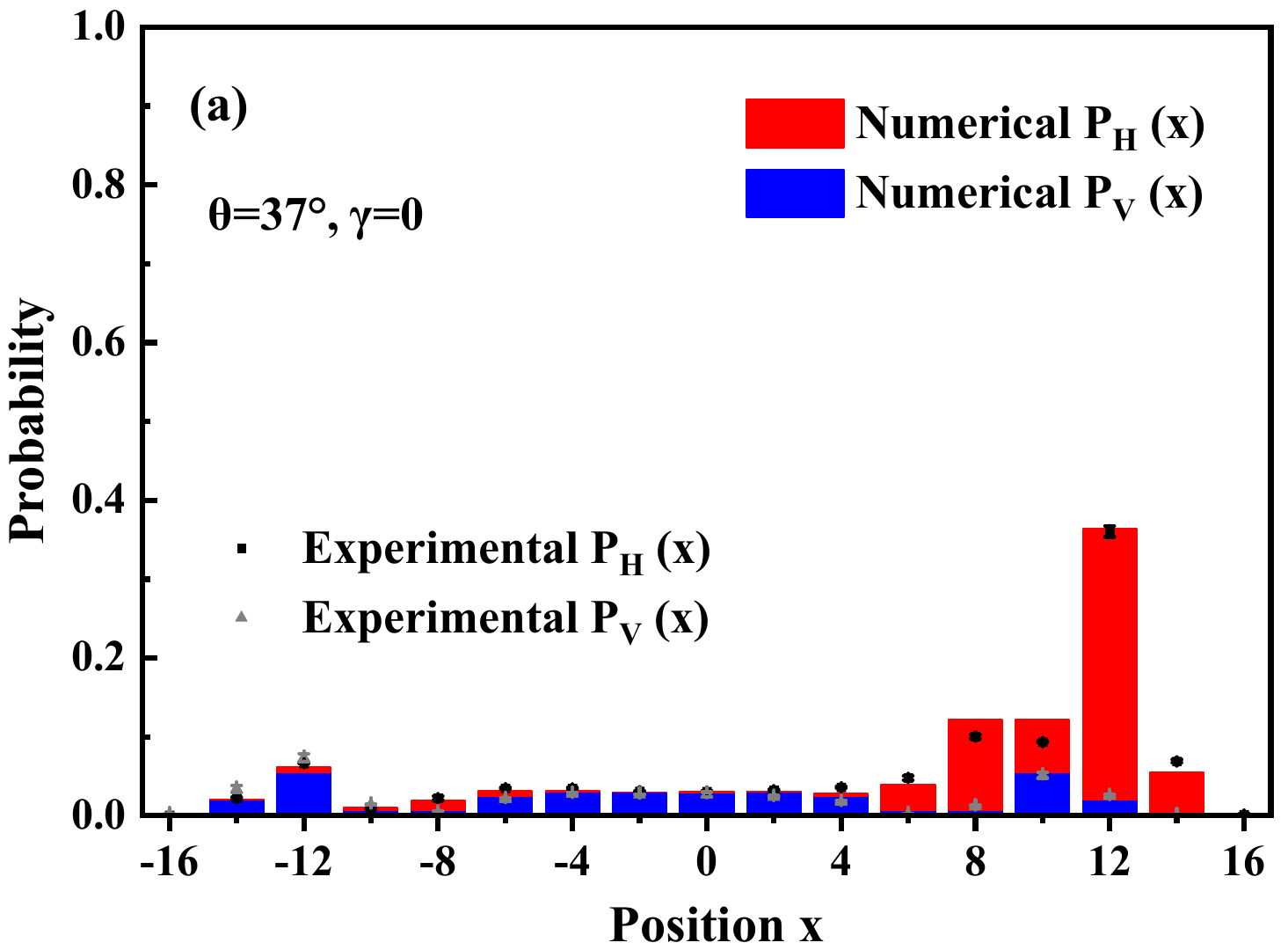} 
	\centering\includegraphics[width=6.5cm]{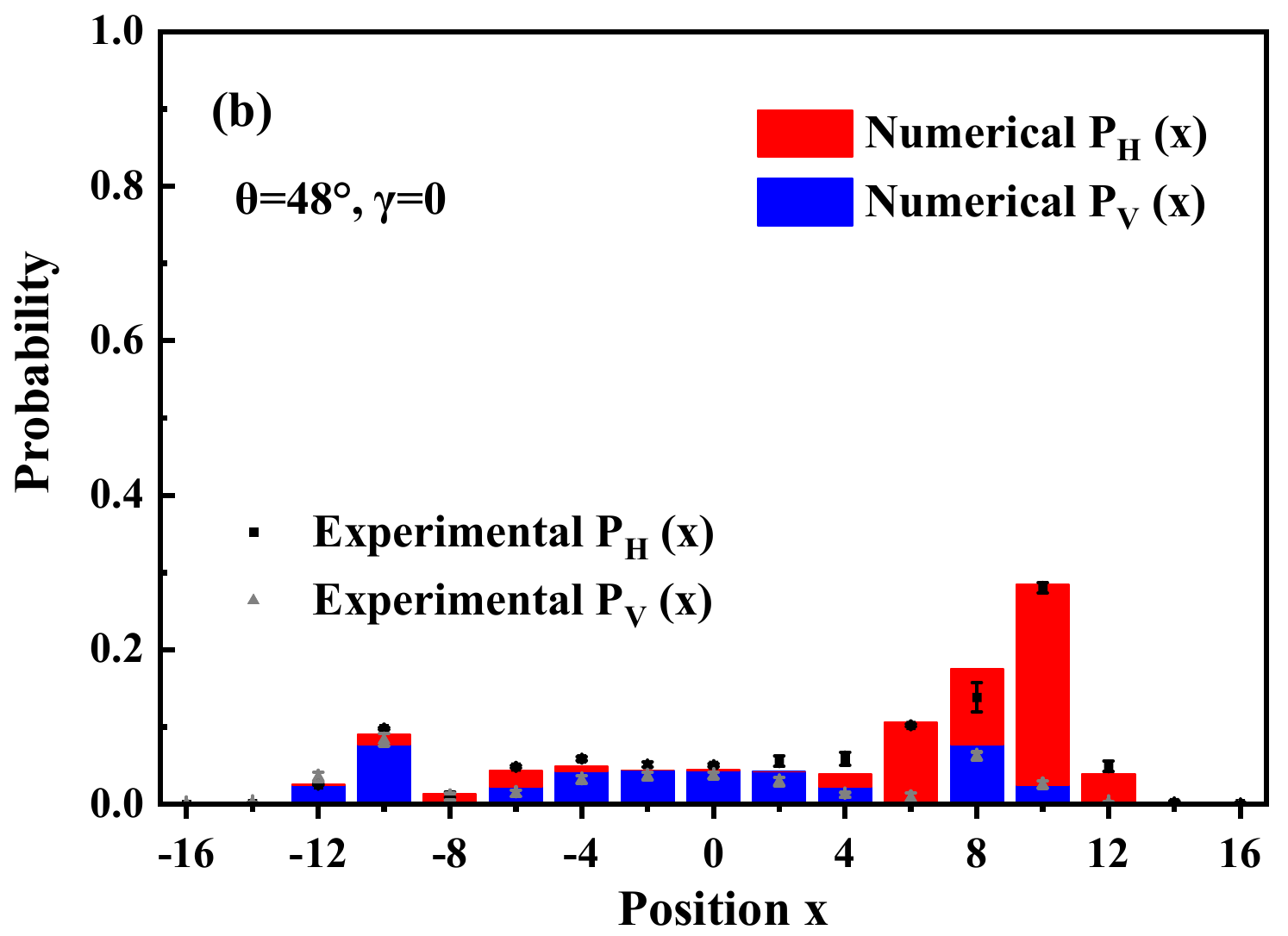}
	\centering\includegraphics[width=6.5cm]{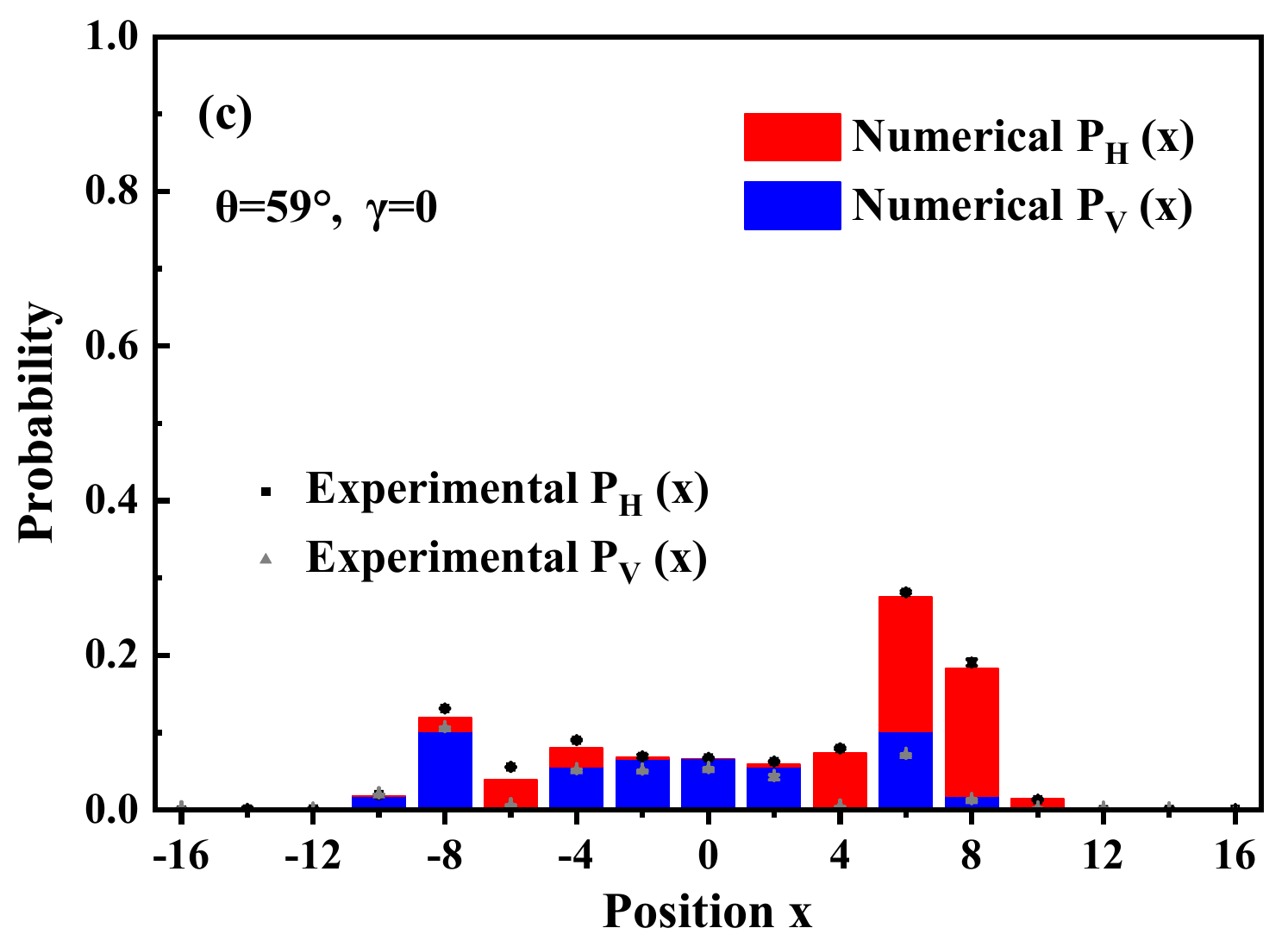} 
	\caption{Experimental and numerical probability distributions of polarized photons in the 16-step quantum walk with the initial state $\left| H \right\rangle \otimes \left| 0 \right\rangle$, as a function of position $x$ for a loss parameter $\gamma = 0$, with (a) $\theta = 37^{\circ}$, (b) $\theta = 48^{\circ}$, and (c) $\theta = 59^{\circ}$. }
\end{figure*}

\begin{figure*}[ht!]
	\centering\includegraphics[width=6.5cm]{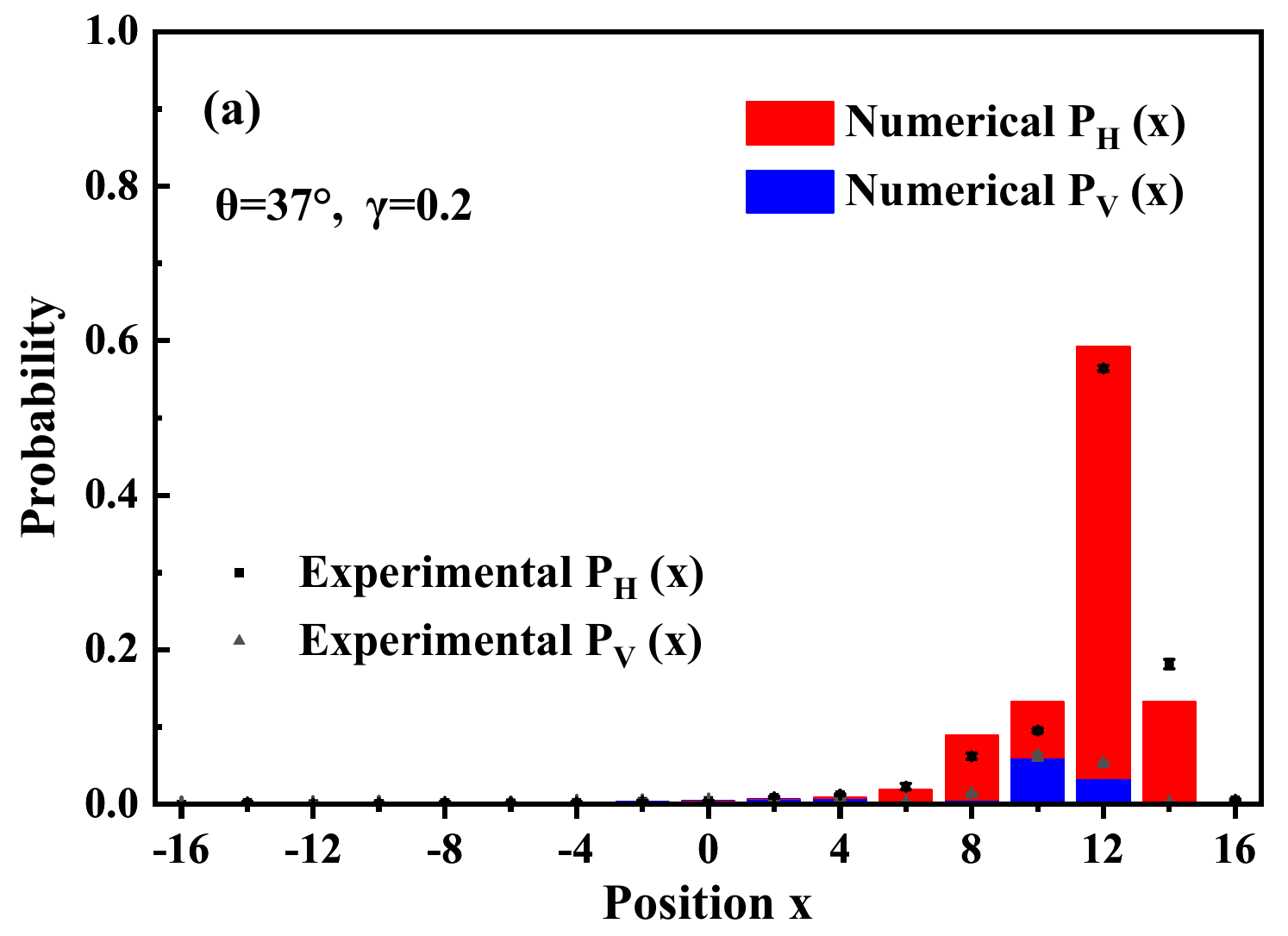} 
	\centering\includegraphics[width=6.5cm]{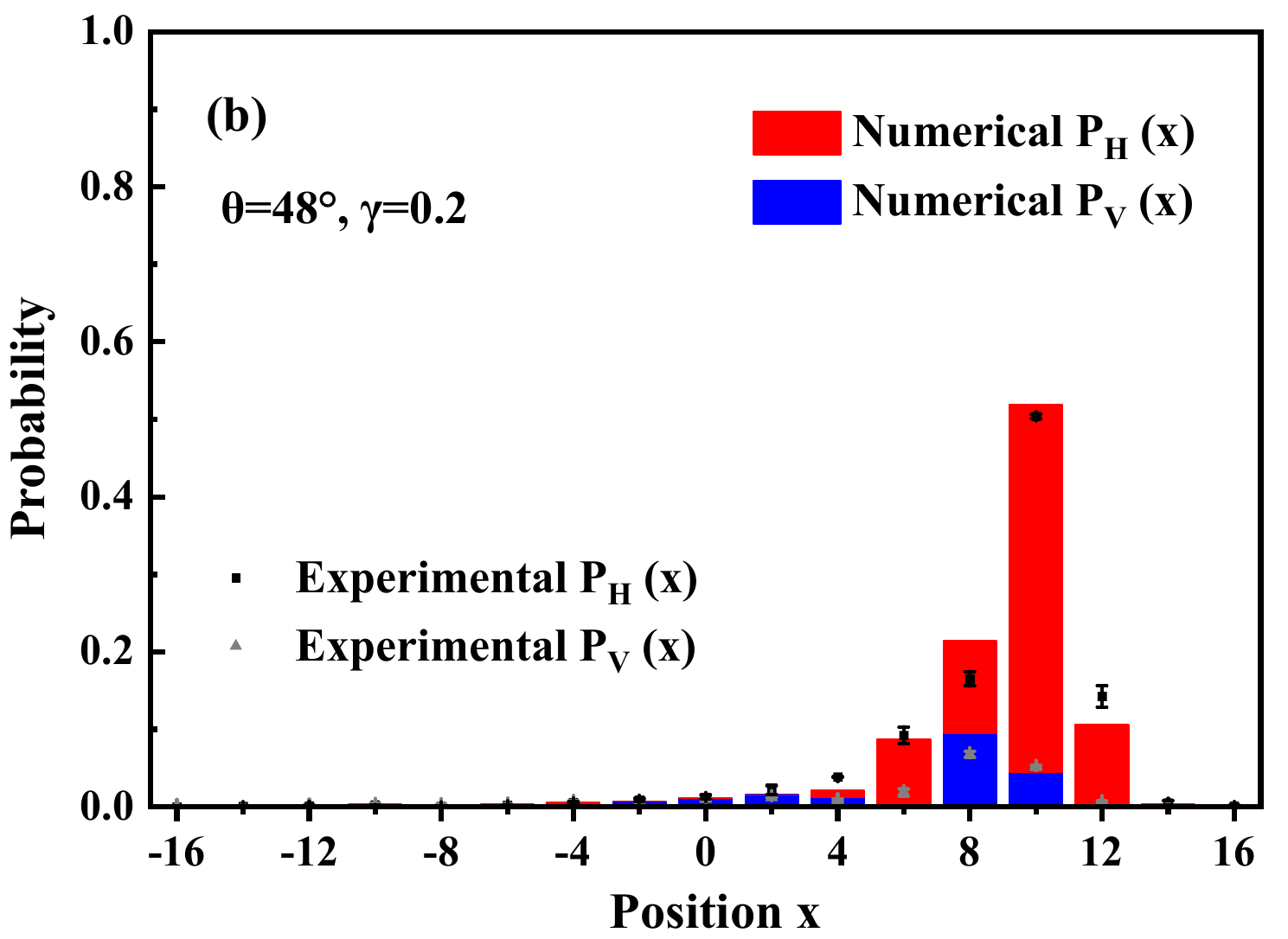}
	\centering\includegraphics[width=6.5cm]{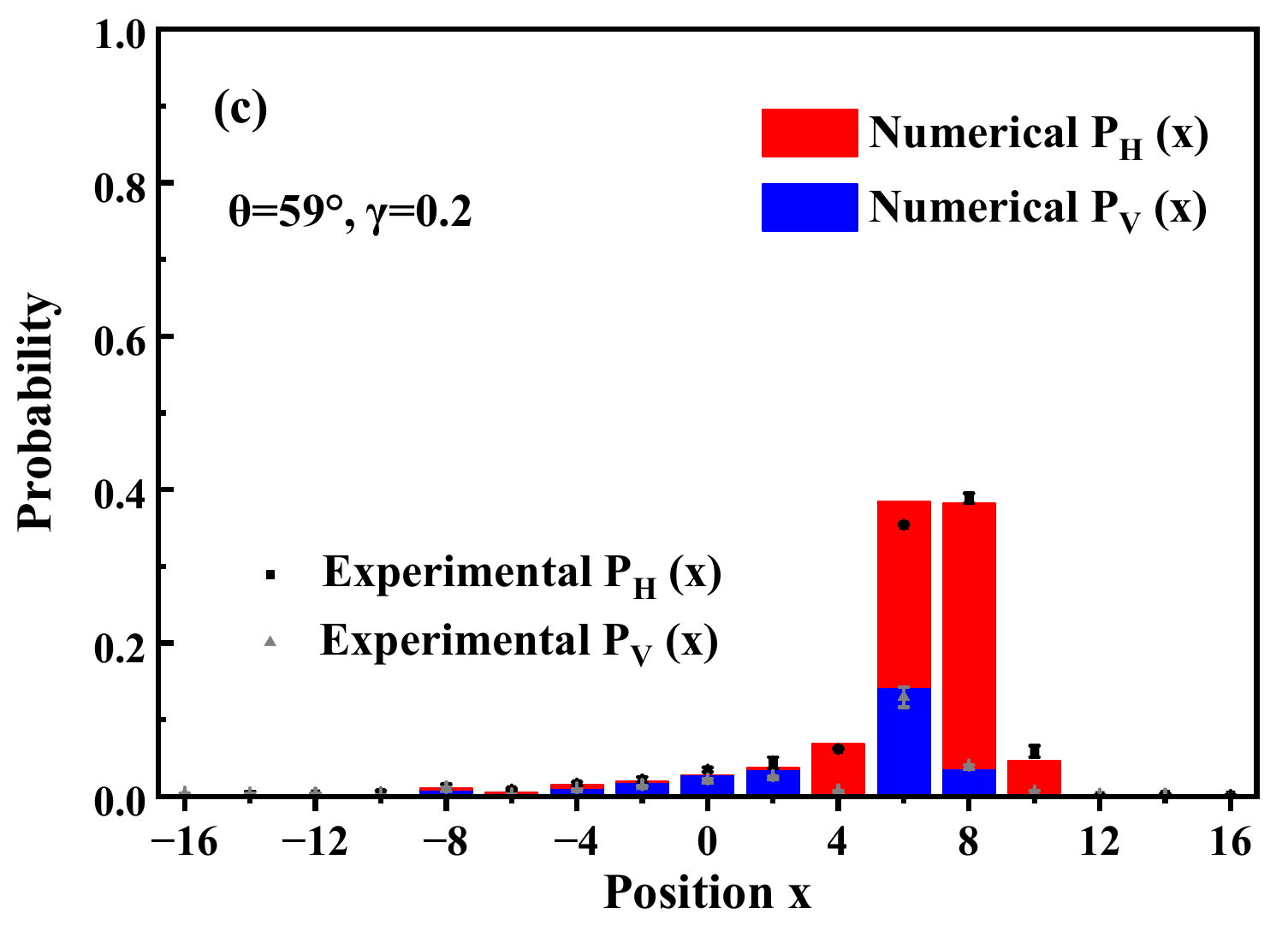} 
	\caption{Experimental and numerical probability distributions of polarized photons in the 16-step quantum walk with the initial state $\left| H \right\rangle \otimes \left| 0 \right\rangle$, as a function of position $x$ for a loss parameter $\gamma = 0.2$, with (a) $\theta = 37^{\circ}$, (b) $\theta = 48^{\circ}$, and (c) $\theta = 59^{\circ}$.}
\end{figure*}
 Fig. 5 (a)-(c) shows the photon probability distribution after 16 steps of the quantum walk with the loss parameter fixed at $\gamma = 0$. The coin parameters are $\theta =37^{\circ }$, $\theta =48^{\circ }$ and $\theta =59^{\circ }$, respectively. The red bars and black squares represent the numerical and experimental results for horizontally polarized photons, while the blue bars and gray triangles correspond to those for vertically polarized photons. For $\theta =37^{\circ }$, the probability distribution reaches its maximum at position $x = 12$ with a value of 0.364. For $\theta =48^{\circ }$, the maximum moves to $x = 10$ with a value of 0.280. For $\theta = 59^{\circ}$, the maximum further moves to $x = 6$ with a value of 0.281. In contrast to the symmetric distribution of a classical random walk, the results show a clear probability enhancement on the right side of the position distribution. This arises because, for the coin parameter $\theta = 0^\circ$ and the initial state $\left| H \right\rangle \otimes \left| 0 \right\rangle$, the coin operator becomes the identity operator and thus preserves the walker's polarization. During the 16-step evolution, the shift operator $\hat{S}$ deterministically moves the photon's position by $x + 1$ at each step. As a result, the photon is located at $x = 16$ with unit probability, while the probability of the photon being at any other position is 0. As the coin parameter increases from $0^{\circ }$, the coin operation gradually transforms the horizontally polarized state $|H\rangle$ into the vertically polarized state $|V\rangle$. After 16 steps of evolution, vertically polarized photons have a nonzero probability of appearing on the left side. As the coin parameter increases from $37^{\circ}$ to $48^{\circ}$ and $59^{\circ}$, the coin operation converts more horizontally polarized photons into vertically polarized photons. This further reduces the maximum probability of photons being on the right side, resulting in a more uniform distribution of photon probabilities on both sides of $x=0$ and thereby enhancing the walker's delocalization.\par

Fig. 6 (a) and (b) show the photon probability distribution after 16 steps of the quantum walk. The loss parameter is fixed at $\gamma = 0.2$. The corresponding coin parameters are $\theta = 37^{\circ}$, $48^{\circ}$, and $59^{\circ}$, respectively. For $\theta =37^{\circ }$, the probability distribution reaches its maximum at position $x = 12$ with an  experimental value of 0.564. For $\theta =48^{\circ }$, the maximum remains at $x = 10$ with a value of 0.505. For $\theta =59^{\circ }$, the maximum moves to $x=8$ with a value of 0.387. Compared with the lossless case ($\gamma = 0$), the maximum photon probability increases significantly after normalization under finite loss parameters. This occurs because, when the coin operator transforms the H-polarized state into the V-polarized state, the loss operator $\hat{L}(\gamma)$ attenuates the V-polarization component. As a result, the probability of the walker being in the $\left|V\right\rangle$ state is reduced. The shift operator $\hat{S}$ moves the walker to $x - 1$ when it is in the $\left|V\right\rangle$ state, thereby suppressing the probability of the walker on the left side. Consequently, under a finite loss parameter, the photon probability is significantly suppressed at the left side, whereas the probability on the right side is significantly enhanced and becomes more localized in the right region. When the coin parameters are set to $59^{\circ }$ and $48^{\circ }$, compared with $37^{\circ }$, the coin operation operators convert more of horizontally polarized states into vertically polarized states. After the shift operator's action, the maximum probability on the right side is reduced, leading to a more uniform photon distribution around $x = 0$. These results confirm that the delocalization of the 16-step quantum walk remains enhanced under finite loss parameters when appropriate coin parameters are chosen.\par

\begin{figure}[ht!]
	\centering\includegraphics[width=6.8cm]{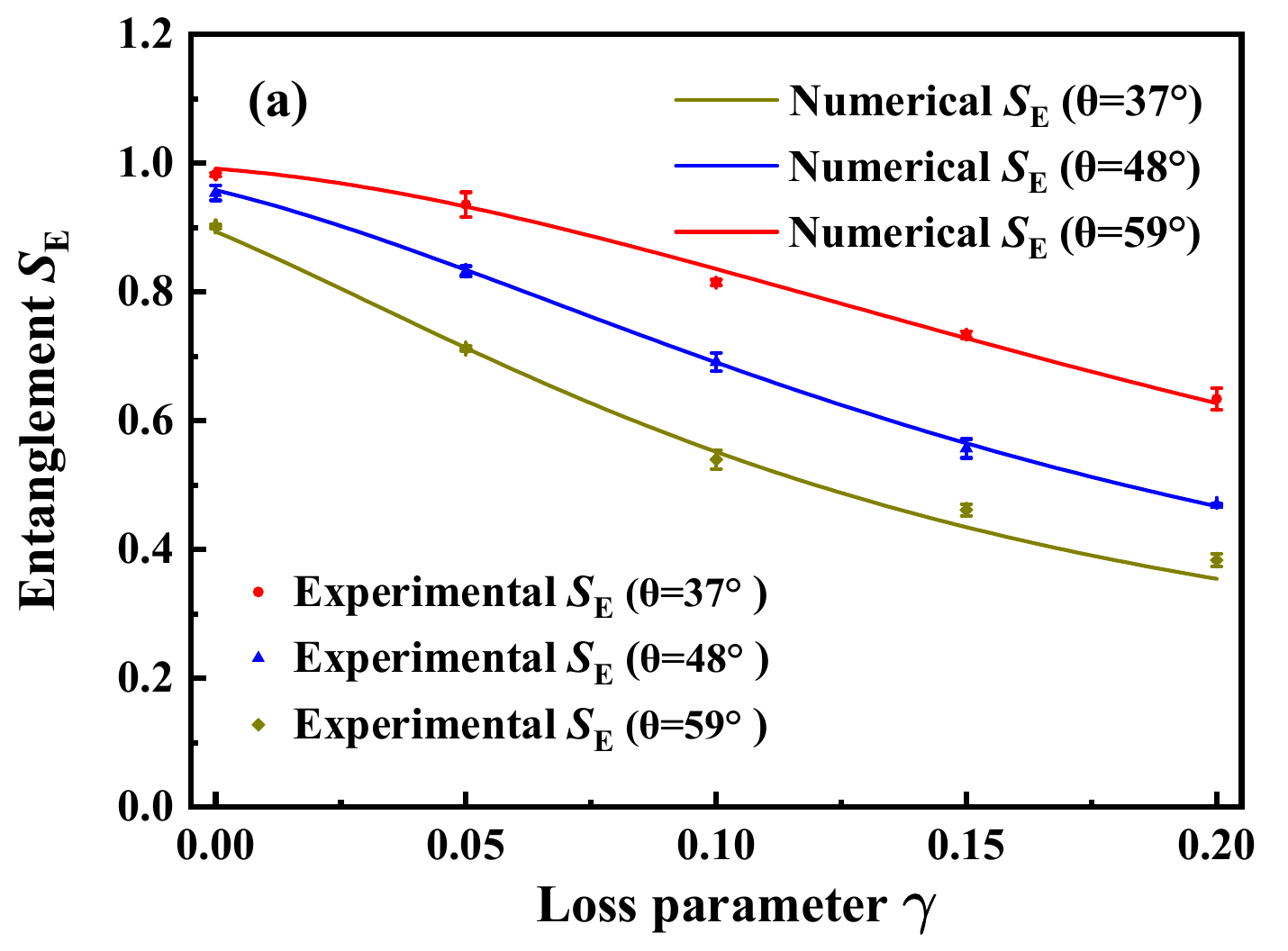} 
	\centering\includegraphics[width=6.5cm]{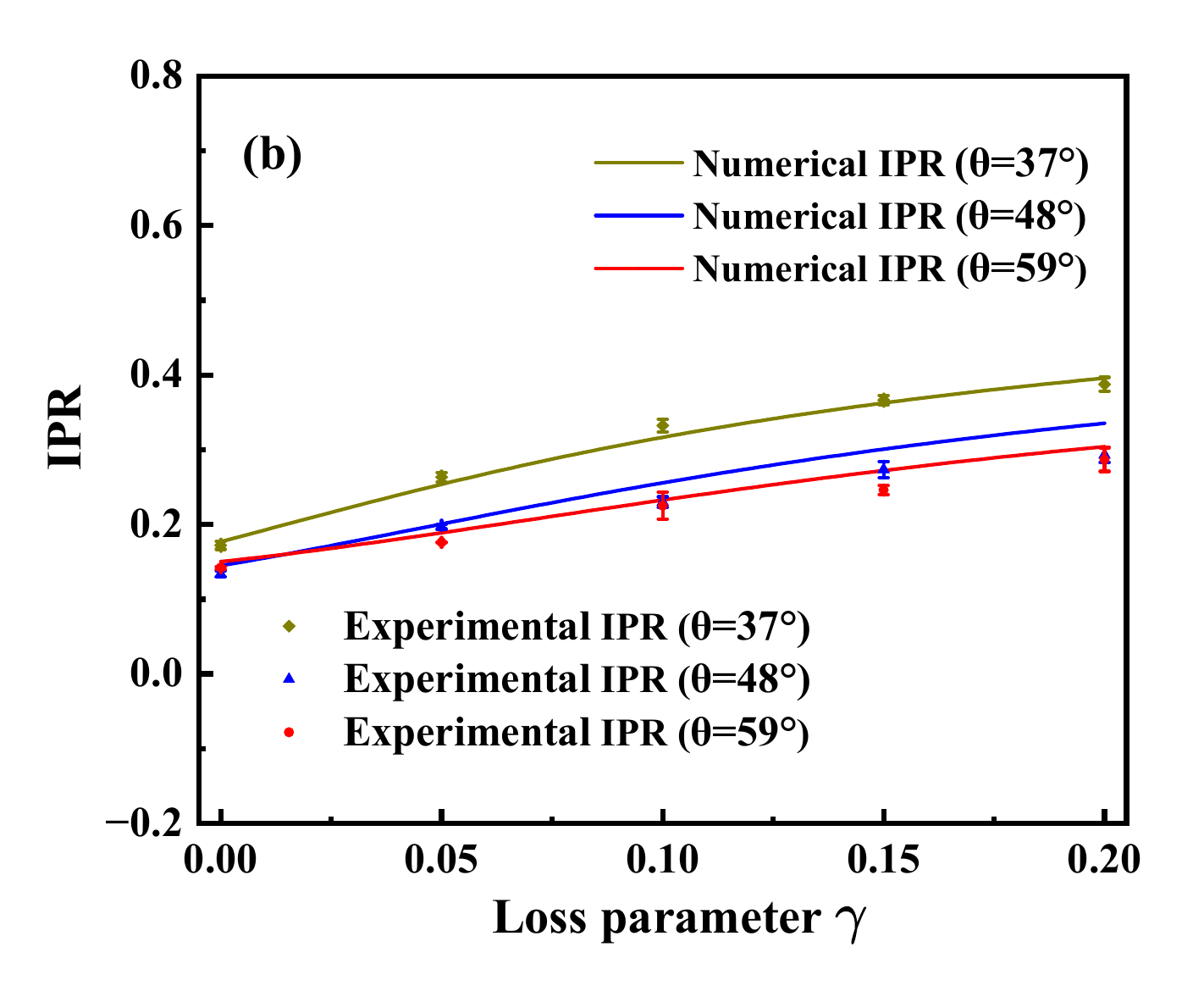} 
	\caption{(a) Coin-position entanglement entropy $S_{E}$ with the asymmetric initial state $\left| H \right\rangle \otimes \left| 0 \right\rangle$ versus loss parameter $\gamma$, for coin parameter $\theta =37^{\circ }$, $\theta =48^{\circ }$, and $\theta =59^{\circ }$. (b) IPR of walker with the asymmetric initial state $\left| H \right\rangle \otimes \left| 0 \right\rangle$ versus the loss parameter, for coin parameter $\theta =37^{\circ }$, $\theta =48^{\circ }$, and $\theta =59^{\circ }$.}
\end{figure}

Fig. 7(a) shows the entanglement entropy $S_{E}$ of the 16-step quantum walk with the asymmetric initial state $\left | H\right \rangle \otimes \left |0 \right \rangle$ versus the loss parameter, based on numerical calculations and experimental data using Eq. (3). The yellow, blue, and red dots correspond to experimental $S_{E}$ values for coin parameters $\theta =37^{\circ }$, $\theta =48^{\circ }$, and $\theta =59^{\circ }$. The solid lines show the numerically predicted values of $S_{E}$. At $\gamma=0$, the measured $S_{E}$ values for $\theta =37^{\circ }$, $\theta =48^{\circ }$ and $\theta =59^{\circ }$ are 0.902 ± 0.003, 0.954 ± 0.011, and 0.986± 0.004, respectively. Thus, for the initial state $\left | H\right \rangle \otimes \left |0 \right \rangle$, the entanglement entropy is significantly enhanced by choosing a coin parameter of $\theta =59^{\circ }$, generating a near-maximally coin-position entangled state. It is also evident that $S_{E}$ decreases with increasing loss parameter $\gamma$, but the decay is significantly slower for $\theta =59^{\circ }$ than for $\theta =48^{\circ }$ and $\theta =37^{\circ }$. Fig. 7(b) shows the IPR of the 16-step quantum walk with the asymmetric initial state $\left | H\right \rangle \otimes \left |0 \right \rangle$ versus the loss parameter, based on numerical calculations and experimental data using Eq. (1). The yellow, blue, and red dots represent the experimental IPR for coin parameters $\theta =37^{\circ }$, $\theta =48^{\circ }$, and $\theta =59^{\circ }$, respectively. The corresponding lines are the numerically predicted IPR. When the loss parameter is set to $\gamma =0$, the experimental IPR values for $\theta =37^{\circ }$, $\theta =48^{\circ }$ and $\theta =59^{\circ }$ are 0.172 ± 0.005, 0.134 ± 0.004, and 0.142± 0.002, respectively. These results, together with Fig. 7(a), confirms that for $\theta =59^{\circ }$, both the entanglement entropy and delocalization of the 16-step quantum walk are enhanced compared to the other cases. Notably, as the loss parameter increases from 0 to 0.2, the IPR for $\theta =59^{\circ }$ and $\theta =48^{\circ }$ consistently remains lower than that for $\theta =37^{\circ }$. Therefore, these results indicate that both the entanglement and the delocalization of the asymmetric DTQW are more robust against polarization-dependent loss within certain ranges of the coin parameter.\par

\section{Conclusion}
In conclusion, this work investigates the delocalization and entanglement properties of walkers in asymmetric quantum walks. The von Neumann entropy and inverse participation ratio are numerically analyzed as functions of the initial state and coin parameters. The results show that, for asymmetric initial states, the regions of coin parameters corresponding to simultaneous enhancement of the walker’s entanglement and delocalization partially overlap, thereby enabling both properties to be improved. Moreover, as the coin parameter increases, the influence of the loss parameter on the walker’s entanglement and delocalization gradually decreases. Based on theoretical analysis, a 16-step asymmetric DTQW was experimentally realized using an optical fiber loop structure. The polarization-resolved photon probability distribution of the 16-step quantum walk was measured for various coin parameters with an asymmetric initial state. When the loss parameter is zero, the photon probability distribution shows a high probability at the right edge and low probability in the center and left regions. For finite asymmetric loss, the photon probability in the left and central regions is strongly suppressed, while the probability at the right edge increases significantly, leading to a more localized distribution toward the right side. Furthermore, for certain coin parameters, we observed a simultaneous enhancement of coin-position entanglement and delocalization for the 16-step quantum walker. As the loss parameter increases, both the entanglement and delocalization of the quantum walk gradually decrease. Interestingly, for specific coin parameters, these properties exhibit improved robustness against polarization-dependent loss in the asymmetric DTQW.\par
Our research shows that DTQWs provide an effective platform for investigating delocalization and hybrid entanglement across multiple degrees of freedom, both of which are critical for quantum information and algorithms. This approach can also be extended to high-dimensional quantum walks to explore advanced phenomena, such as multi-particle hybrid entanglement and high-dimensional non-Hermitian effects. \par

This work was supported by the Scientific and Technological Research Program of the Education Department of Hubei Province under Grant No. B2023139 and No. Q20222502, the Innovation Development Union Fund of Huangshi city under Grant No. 2024AFD010, and the Innovation and Entrepreneurship Training Program of Hubei Province under Grant No. S202410513098.

\bibliography{sample}
\end{document}